\documentclass[12pt]{iopart}

\usepackage{graphicx}
\usepackage[utf8]{inputenc} 
\usepackage[T1]{fontenc}    
\usepackage{xcolor}
\definecolor{darkblue}{RGB}{46,48,147}
\usepackage[colorlinks=true,
            linkcolor=darkblue,
            urlcolor=darkblue,
            citecolor=darkblue]{hyperref}
\usepackage{url}            
\usepackage{booktabs}       
\usepackage{amsfonts}       
\usepackage{nicefrac}       
\usepackage{microtype}      
\usepackage{multirow}
\usepackage{xspace}
\usepackage{amssymb}
\usepackage{todonotes}
\usepackage{cite}

\graphicspath{{./}}

\begin{document}

\title{Source-Agnostic Gravitational-Wave Detection with Recurrent Autoencoders}

\author{Eric A. Moreno~\footnote{Previously at California Institute of Technology.}}
\address{Massachusetts Institute of Technology, Cambridge, Massachusetts 02139, USA}

\author{Bartlomiej Borzyszkowski~\footnote{Also at CERN and Intel Technology Gdansk, Poland.}}
\address{Gdansk University of Technology, Narutowicza 11/12, 80-233, Gdansk, Poland}

\author{Maurizio Pierini}
\address{European Organization for Nuclear Research, 1211 Geneva 23, Switzerland}

\author{Jean-Roch Vlimant, Maria Spiropulu}
\address{California Institute of Technology, Pasadena, California 91125, USA\vspace{\baselineskip}}

\ead{maurizio.pierini@cern.ch}
\vspace{10pt}
\begin{indented}
\item[]July 2021
\end{indented}

\begin{abstract}
We present an application of anomaly detection techniques based on deep recurrent autoencoders to the problem of detecting gravitational wave signals in laser interferometers. Trained on noise data, this class of algorithms could detect signals using an unsupervised strategy, i.e., without targeting a specific kind of source. We develop a custom architecture to analyze the data from two interferometers. We compare the obtained performance to that obtained with other autoencoder architectures and with a convolutional classifier. The unsupervised nature of the proposed strategy comes with a cost in terms of accuracy, when compared to more traditional supervised techniques. On the other hand, there is a qualitative gain in generalizing the experimental sensitivity beyond the ensemble of pre-computed signal templates. The recurrent autoencoder outperforms other autoencoders based on different architectures. The class of recurrent autoencoders presented in this paper could complement the search strategy employed for gravitational wave detection and extend the discovery reach of the ongoing detection campaigns.
\end{abstract}

\vspace{2pc}

\maketitle
%
%

\section{Introduction}
\label{sec:intro}

The detection of gravitational waves (GW) from stellar binaries such as black hole and neutron star mergers have ushered in a new era of analyzing the Universe. Operating together, the Laser Interferometer Gravitational-wave Observatory (LIGO) \cite{TheLIGOScientific:2014jea} and the 
Virgo Interferometer~\cite{TheVirgo:2014hva}
can peer into deep space giving astronomers the ability to uncover and localize stellar processes through their gravitational signature. The first observation of a binary black hole merger (GW150914)~\cite{PhysRevLett.116.061102} has given way to a plethora of GW events, and notably to the observation of intermediate-size black holes~\cite{Abbott:2017vtc} and neutron star mergers~\cite{TheLIGOScientific:2017qsa}, an event that marked the beginning of the multi-messenger astronomy era~\cite{GBM:2017lvd}.

Instrumental on the software side of these observations are the algorithms which identify the faint GW signals in an environment characterized by overwhelming classical and quantum noise. The most sensitive detection algorithm, Matched Filtering (MF)~\cite{Allen:2005fk}, consists of matching incoming data with templates of simulated GW shapes, covering the parameter space of binary masses ~\cite{PhysRevD.44.3819, PhysRevD.53.3033, PhysRevD.53.6749, PhysRevD.76.102004, Smith:2016qas}, which is then used to identify the signal. At the same time, they offer an estimate of the astrophysical parameters associated to the detected GW event, such as the nature and mass of the two merging objects. This method was extremely functional to the success of the LIGO and VIRGO observation campaigns. On the other hand, by relying on pre-computed templates, it could lead to missing events for which a template is not available. This could be an event originating from computationally prohibitive conditions~\cite{PhysRevD.93.042004, PhysRevD.95.024038, Huerta:2013qb, Huerta:2014eca}, or even some sort of unforeseen GW source. To compensate for the limitations implicit in these assumptions, the data processing pipelines of LIGO and VIRGO also included  the coherent WaveBurst~\cite{Drago:2020kic}, an alternative strategy designed to deal with unmodeled GW sources.

Detecting GWs is certainly one of the hardest challenges faced in fundamental science in the recent years. Given how weak a GW signal is when compared to typical noise levels, it is natural to look at Machine Learning (ML), and especially to Deep Learning (DL), in order to improve a signal detection capability. For instance, an early attempt at multivariate classification with random forests~\cite{Baker:2014eba} are furthered by Refs.~\cite{George:2016hay, George:2017pmj,Gabbard:2017lja, Jadhav:2020oyt} which discuss how Convolutional networks~\cite{fukushima:neocognitronbc, LeCun1999} can be trained to extract a variety of GW signals from highly noisy data. This is an example of a supervised classifier, i.e., a classifier trained to separate different populations in data (e.g., signal vs. noise) by matching a given set of ground-truth labels . While classifiers could certainly contribute to enhance the current state-of-the-art detection capability, they rely on pre-defined signal much like the MF technique. In other words, they are designed to possibly improve the detection accuracy but they are not necessarily going to extend the detector sensitivity to exotic signals outside the portfolio of pre-simulated templates. In fact, these networks are typically trained on labelled data from simulation, so even in this case the capability of simulating a given signal is an underlying requirement. On the other hand, GW detection comes with the need of going beyond signal signatures that can be emulated. Ref.~\cite{McGinn:2021jqg} discusses how to generate “unmodeled” waveforms, which could then be used to train a supervised algorithm without the use of templates.

Besides the search for exotic sources, model independent strategies could be useful to deal with practical issues such as glitch detection~\cite{Colgan:2019lyo}.

In this paper, we investigate the possibility of  rephrasing the problem of GW detection as an anomaly detection task. 
By anomaly detection we mean the use of a one-class classifier, in this case a Deep Autoencoder (AE)~\cite{AE}, to identify outlier populations in an unlabelled dataset. An autoencoder is a compression-decompression algorithm that is trained to map a given set of inputs into itself, by first compressing the input into a point in a learned latent space (encoding) and then reconstructing it from the encoded information (decoding).

Once trained on standard events, the autoencoder might fail to reconstruct samples of different kind (the anomalies). Any input-to-output distance measurement can then be used to identify these anomalies. Under the assumption that these anomalies are rare, one can directly train these AEs on data, looking for the set of AE parameter values that minimize the difference between the input and the output, using some distance {\cal D} as a loss function. By taking as a reference the distribution of {\cal D} on data, one can label anomalies as the outlier events of this distribution. This very same approach was recently discussed in Ref.~\cite{Morawski:2021kxv}, where the discovery reach of Convolutional autoencoders for GW detection is investigated. In this work, we consider two recurrent AE architectures: Long-short memory networks (LSTMs)~\cite{lstm}, and Gated Recurrent Units (GRUs)~\cite{DBLP:journals/corr/ChungGCB14}). For comparison, we consider alternative AE architectures: dense (i.e., fully connected) neural networks (DNNs) and Convolutional Neural Networks (CNNs). 

The main advantage of this unsupervised strategy is that one algorithm is potentially sensitive to multiple signal typologies. On the other hand, this gain in flexibility is typically followed by a loss in accuracy. For a specific signal, an algorithm trained with an unsupervised procedure on unlabeled data is typically less accurate than a supervised classifier trained on labeled data. 

The proposed strategy comes with another remarkable advantage: under the assumption that anomalies are rare instances in the processed data, autoencoders can be trained directly on data, without relying on signal or noise simulation. Instead, supervised algorithms require labels which, in the case of rare signals like those considered in this study, are obtained using synthetic data (e.g., from Monte Carlo simulation). Assuming this training could happen in real time, the AE could adapt to changing experimental conditions and limit the occurrence of false claims.

This paper is organized as follows: related works describing ML approaches to GW detection are briefly discussed in Section~\ref{sec:relatedWorks}. The dataset utilized for this study is described in Section~\ref{sec:data}. The autoencoder architectures are described in Section~\ref{sec:AE}, together with the corresponding classifiers used to benchmark performance. Results are presented in Section~\ref{sec:results}. Conclusions are given in Section~\ref{sec:conclusions}.

\section{Related work}
\label{sec:relatedWorks}

Besides MFs, 

In addition, Bayesian inference libraries~\cite{Veitch:2014wba,Ashton_2019,10.1093/mnras/staa2850} have been created to estimate the properties of a gravitational-wave source. Unsupervised anomaly detection of transients  has also been proposed using a temporal version of k-nearest neighbors (kNN)~\cite{Benko:2020syv}. 

Most DL approaches for GW classification~\cite{Baker:2014eba,George:2016hay, Kapadia:2017fhb,George:2017pmj,Gabbard:2017lja,Miller:2019jtp,Jadhav:2020oyt,Huerta:2020xyq} involve supervised learning techniques, which typically provide competitive accuracy by exploiting network non-linearity and the information provided by ground-truth labels. By construction, these methods rely on a realistic simulation of the signal induced by a specific kind of source, which is assumed upfront.

Principal Component Analysis~\cite{doi:10.1080/14786440109462720} (PCA), which performs a linear orthogonal transformation of a set of (possibly correlated) variables into a set of linearly
uncorrelated variables, has also been introduced for transient detection~\cite{Powell_2015, Powell_2017}. This can give a quick characterization of the intrinsic properties of a data sample. The DL methods discussed in this paper are generalizations of this approach to include nonlinear compression.

Boosting Neural Networks~\cite{10.5555/1294154.1294155}, which use a combination of unsupervised and supervised learning techniques, have also been used for GW classification~\cite{PhysRevD.95.104059}. This method performs an unsupervised hierarchical clustering on the incoming data to identify possible groups and a supervised Bayesian classifier to do the final classification. BNNs can classify a number of different injections including Gaussian, Ringdowns, Supernovae, white noise bursts, mergers, etc. Importantly, the architecture requires sufficient statistics to cluster in an unsupervised manner, which makes it not ideal for the processes that we focus on in this paper. 
 
Finally, CNNs have achieved high accuracy while having the added benefit of parameter estimation to infer useful parameters of the GWs such as masses and spins of the binary merger components~\cite{George:2016hay, George:2017pmj}. In a similar approach to the one discussed in this paper, Ref.~\cite{Morawski:2021kxv} discusses the use of CNN autoencoders in order to classify GWs. For comparison, the same architecture as specified in \cite{Morawski:2021kxv} is implemented in Section~\ref{sec:AE} alongside the recurrent autoencoders. 

\section{Data samples}
\label{sec:data}

\begin{figure}[t!]
   \centering
   \includegraphics[width=0.9\columnwidth]{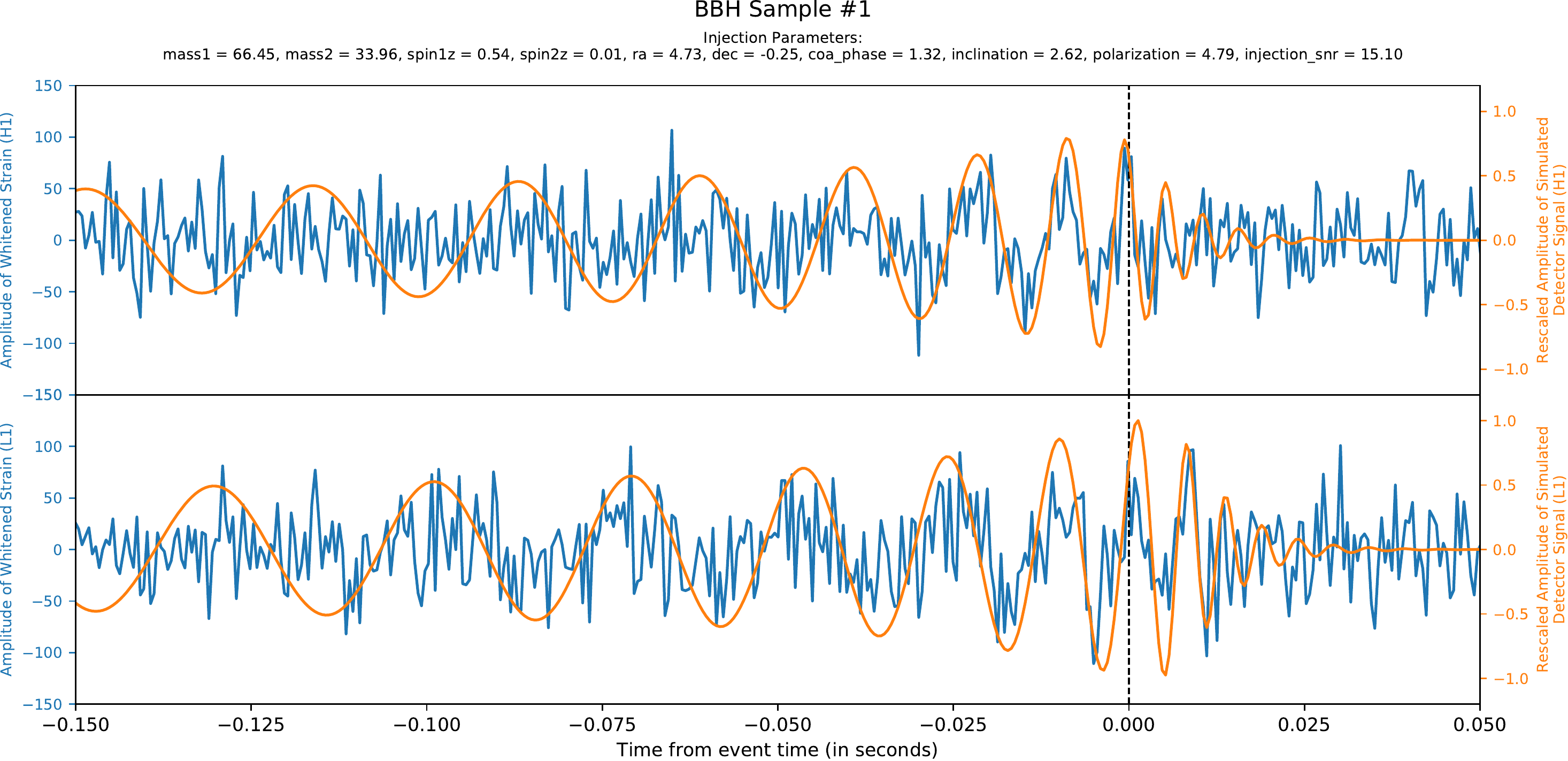}
   \newline
   
   \includegraphics[width=0.9\columnwidth]{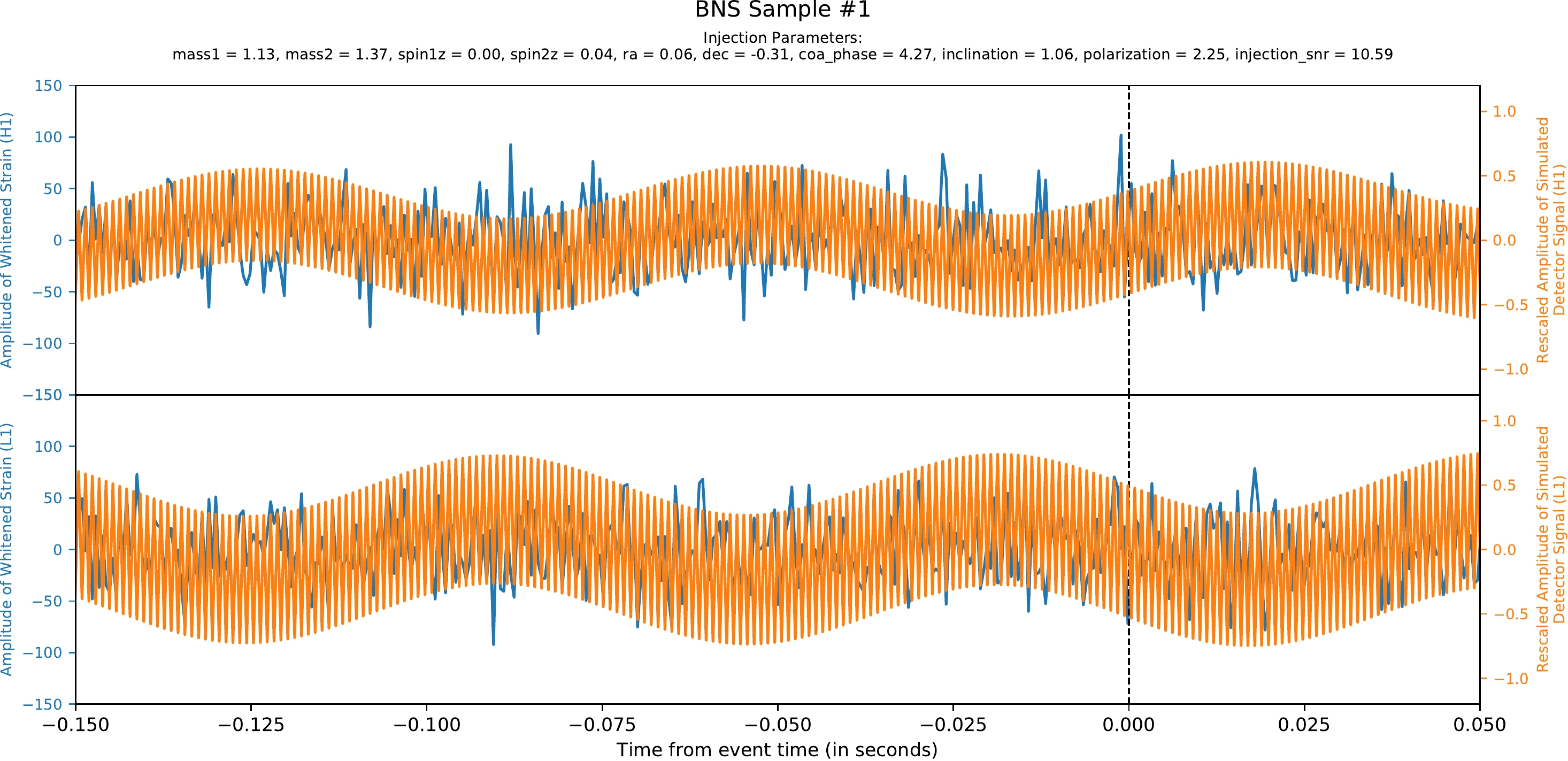}
    \caption{Resulting strains from simulated gravitational waves on the Livingston (L1) and Hanford (H1) detectors from the coalescence of a 66 M\textsubscript{\(\odot\)} and 34 M\textsubscript{\(\odot\)} black holes (top) and 1.1 M\textsubscript{\(\odot\)} and 1.4 M\textsubscript{\(\odot\)} neutron stars (bottom), obtained using the GGWD package~\cite{GGWD}. The amplitude of the GW signal, which has been arbitrarily scaled to be more visible, is illustrated in orange and the strain on the detector including this signal is illustrated in blue. As the amplitudes are arbitrarily scaled in this figure, it is important to note that binaries with lower mass will have longer waveforms and, for the same integrated SNR, will have a smaller instantaneous amplitude.}
    \label{fig:GW_sample}
\end{figure}

This study is performed on a sample of synthetic data, generated for this study and available on Zenodo~\cite{eric_moreno_2021_5772814, eric_moreno_2021_5773513}. Data are generated using publicly available LIGO software suites~\cite{lalsuite, alex_nitz_2020_3993665} and processed using the GGWD library~\cite{GGWD}. 

Noise events occur when no signal is overlapped to the detector noise. Detector noise is generated with an aLIGOZeroDetHighPower \cite{lalsuite, PSD} power spectral density (PSD), which is computed as a function of the length of noise, time step of the noise, and noise weighting to color the noise. This is done using PyCBC ~\cite{alex_nitz_2020_3993665}. This approach to simulated data generation ignores glitches, blips, and other transient sources of detector noise.~\footnote{For this reason, the considered AEs could be re-purposed for anomaly detection algorithms to identify detector glitches. The main difference between these kind of glitch anomalies and those of astronomical significance would be the lack of coincidence of anomalies across different detectors, a clear indication of an anomalous signal of astrophysical origin.}

In absence of exotic signal sources, we use traditional GW signals to assess the detection performance. We consider two kinds of GW sources from binary mergers: Binary black hole (BBH) and Binary neutron star (BNS). Signal events are generated simulating GW production from compact binary coalescences using PyCBC~\cite{alex_nitz_2020_3993665}, which itself uses algorithms from LIGO's LAL Suite~\cite{lalsuite}. Signal event containing GWs were created overlaying simulated GWs on top of detector noise. This provides an analogous situation to a real GW, in which the strain from the incoming wave is recorded in combination with the normal detector noise. 




The dataset consists of {400,000} noise samples. Each sample corresponds to 8 seconds of data, sampled at 2048 Hz. We consider a LIGO-like experimental setup, with two detectors (L1 and H1) taking data simultaneously at different locations. The simulation also includes a difference in GW time-of-arrival at each detector due to light travel time between them, which is significant at a sampling rate of 2048 Hz. Within a signal 8-second event, the PSD estimate is used to calculate the network optimal matched filtering signal-to-noise ratio (NOMF-SNR), which is then used to scale the added injection to form a uniform SNR sample distribution. For each detector, a sample is represented as a one-dimensional array with 16,384 entries.

Figure~\ref{fig:GW_sample} shows two of the simulated signal events: a coalescence of two black holes, with masses set to 66 and 34 solar masses (M\textsubscript{\(\odot\)}), and a coalescence of two neutron starts, with masses set to 1.1 and 1.4 solar masses (M\textsubscript{\(\odot\)}), detected by the L1 and H1 detectors. The generation parameters are listed in the figure: the spin of the two black holes, right ascension, declination, coalescence phase, inclination, polarization, and signal-to-noise ratio~\cite{GGWD}. In absence of any source of noise, the signal would appear as shown by the orange line. Once the noise is added, the detectable signal corresponds to the blue line. 

This dataset is split in three parts: 256,000 training samples ($64 \%$), 64,000 validation samples ($16\%$), and 80,000 test samples ($20 \%$). The training and validation datasets are used in the optimal-parameter learning process, while the test dataset is used to assess the algorithm performance after training, together with the signal samples.  
\newline

\noindent{}
The BBH sample is generated with the following parameters and priors:
\begin{itemize}
    \item SEOBNRv4~\cite{Bohe:2016gbl} Approximant.
    \item Masses independently and uniformly varied within [10, 80]  M\textsubscript{\(\odot\)}.
    \item Spins independently and uniformly varied within [0, 0.998]. 
    \item Injection network signal-to-noise ratio (SNR) uniformly varied within [5, 20] for full 8-second event.
    \item Coalescence phase uniformly varied within [0, 2$\pi$].
    \item Inclination varied with Sine prior from [0, $\pi$].
    \item Right Ascension sampled uniformly from [0, 2$\pi$] using uniform\_sky prior~\cite{alex_nitz_2020_3993665}. 
    \item Declination sampled uniformly from [-$\pi$/2, $\pi/$2] using uniform\_sky prior~\cite{alex_nitz_2020_3993665}.
    \item Polarization sampled uniformly from [0, 2$\pi$].
\end{itemize}
The BNS sample is generated with the following parameters and priors:
\begin{itemize}
    \item IMRPhenomDNRTidal\_v2~\cite{Abbott:2018wiz} Approximant.
    \item Masses independently and uniformly varied within [1.1, 2.1] M\textsubscript{\(\odot\)}  \cite{_zel_2012}.
    \item Spins independently and uniformly varied within [0, 0.05] \cite{Mandel:2009nx}.
    \item Injection network signal-to-noise ratio (SNR) uniformly varied within [5, 20] for full 8-second event.
    \item Coalescence phase uniformly varied within [0, 2$\pi$].
    \item Inclination varied with Sine prior from [0, $\pi$].
     \item Right Ascension sampled uniformly from [0, 2$\pi$] using uniform\_sky prior~\cite{alex_nitz_2020_3993665}. 
    \item Declination sampled uniformly from [-$\pi$/2, $\pi/$2] using uniform\_sky prior~\cite{alex_nitz_2020_3993665}.
    \item Polarization sampled uniformly from [0, 2$\pi$].
\end{itemize}

Data are whitened with a Fast Fourier Transform integration length of 4 seconds and a duration of the time-domain Finite Impulse Filter whitening filter of 4 seconds to remove the underlying correlation in the data~\cite{Cuoco:2000gv}. Then, band-pass filtering was applied to remove high frequency (above 2048 Hz) and low frequency (below 30 Hz) components from the data. Doing so, background from outside the current interferometer sensitivity range is discarded. 
To facilitate the data processing and learning by the network, the data are scaled absolutely to a [0, 1] range. Each 8 second event is then cropped to 2.5 sec around the GW event, with the GW happening at a random time after the beginning of the time window. The GW arrival is uniformly sampled in a [1, 2] sec interval. This choice allows us to take into account the appropriate time for the signal ring-up/ring-down. This is done to assure that the model classifiers are not biased to a certain time period within the event, which would occur if the simulated data has GWs appearing at only a single time within the event window. 

\section{Network architectures}
\label{sec:AE}

Autoencoders are algorithms that project an input sample $x \in {\cal X}$ to its encoded projection $z$ in a latent space ${\cal Z}$, typically of lower dimension than the input space. The  encoded projection $z$ is then decoded to a reconstructed $\hat{x} \in {\cal X}$. The network parameters determine how $x$ is projected to $z$ and then back to $\hat{x}$. Their values are fixed minimizing some input-to-output distance, used as a loss function in the network training. In this study, we consider the mean-squared error (MSE) between each element of the input array and the corresponding output. 
We consider several network architectures, all structured according to a common scheme with the decoder mirroring as close as possible the encoder architecture. Three specific architectures are introduced, with CNN, LSTM, or GRU layers. 

As a comparison to recurrent layers, we implement a 1D CNN AE similar to Ref.~\cite{Morawski:2021kxv} with an input of 25 one-dimensional windows of shape (1024, 1) slid around a 2.5-second interval sampled at 2048 Hz. Each of these 25 inputs are individually inputted into the AE, producing 25 reconstructed steps (and losses) to work with in post-processing analysis. The encoder consists of two one-dimensional convolutional layers with filters of size [256, 128] and kernel of size 3, coupled with a maxpool layer of size 2. The decoder mirrors this architecture with an upsampling layer of size 2, and two one-dimensional convolutional layers with filter size [256, 1] and kernel of size 3. In addition, DNN AEs with number of nodes {100, 50, 10} and {10, 50, 100} were attempted but yielded worse results. This is expected as DNN AEs are not specialized in time-series data, unlike recurrent architectures.

\begin{figure}[t!]
  \centering
  \includegraphics[width=0.9\columnwidth]{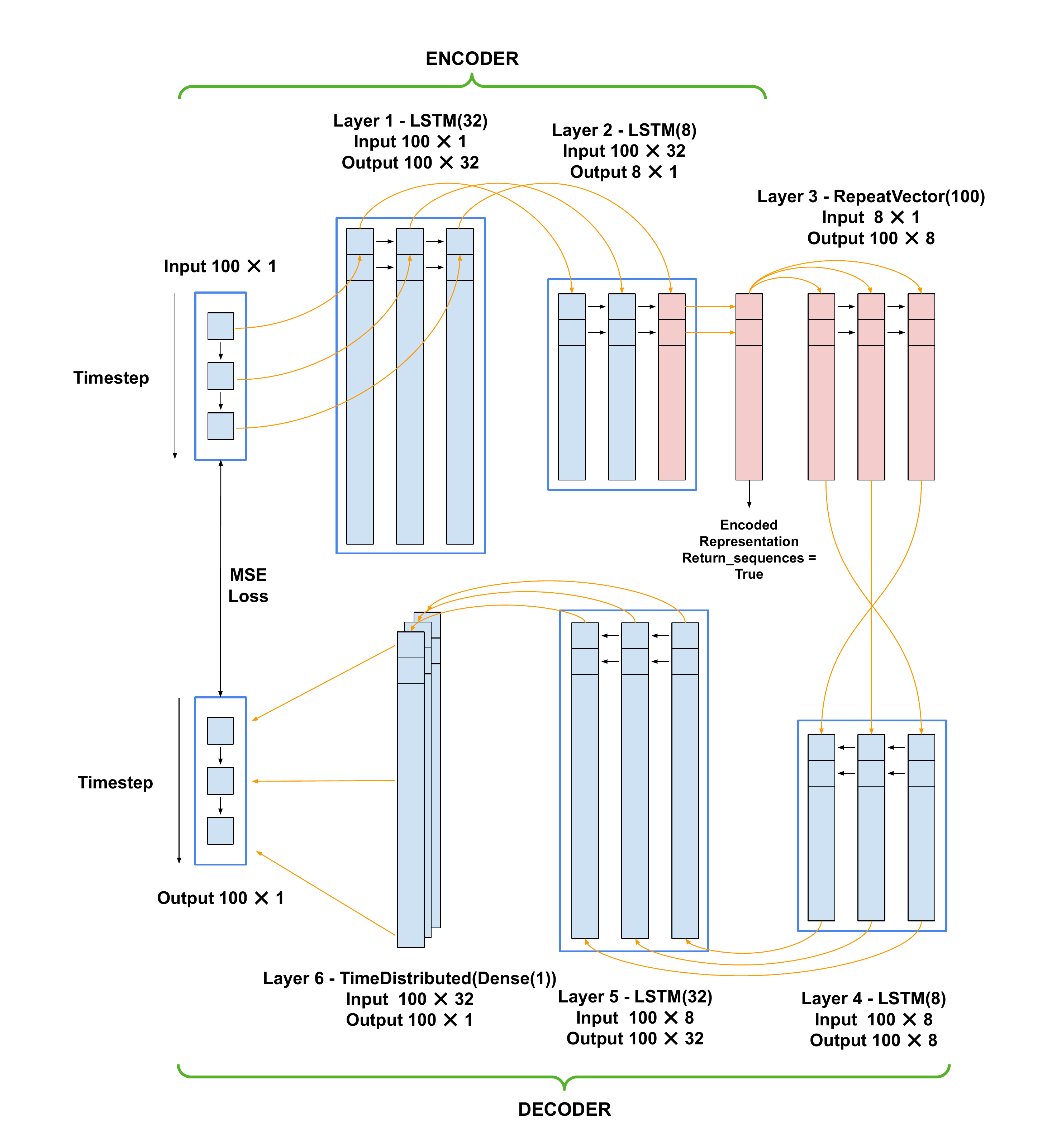}
  \caption{Graphical illustration of LSTM autoencoder. The general encoder structure still remains, and the latent space is created by setting return\_sequences=False in the LSTM cell, passing through only the final time step of the LSTM. This latent space is then repeated using a RepeatVector, forcing the encoder to create some important latent representation as its final output which can then be sent to the decoder. A similar workflow can be established for a GRU autoencoder or any other recurrent autoencoder. 
  \label{fig:LSTM-AE}}
\end{figure}

The LSTM and GRU networks function similarly, utilizing LSTM and GRU cells instead of simple CNN/DNN nodes. The latent space bottleneck in this representation is created by instructing the LSTM (or GRU) cells to only return their final state in the encoding phase and then repeating that final state as a vector which can then be input to the decoder. The input to these recurrent architectures are 256 one-dimensional arrays of shape (100, 1) slid around a 2.5-second interval sampled at 2048 Hz. Each of these 256 inputs are individually inputted into the AE, producing 256 reconstructed steps (and losses) to work with in post-processing analysis. The encoder consists of 2 layers, with number of units 32 and 8, which produces a latent representation by only outputting the final LSTM state vector on the bottleneck layer. The decoder consists of two layers with 8, 32 nodes, which is then multiplied by a final temporal slice of a dense layer, yielding the same dimensions as the input representation.  For illustration, the architecture of the LSTM model is shown in Fig.~\ref{fig:LSTM-AE}.

\begin{figure}[t!]
    \centering
    \includegraphics[width=\columnwidth]{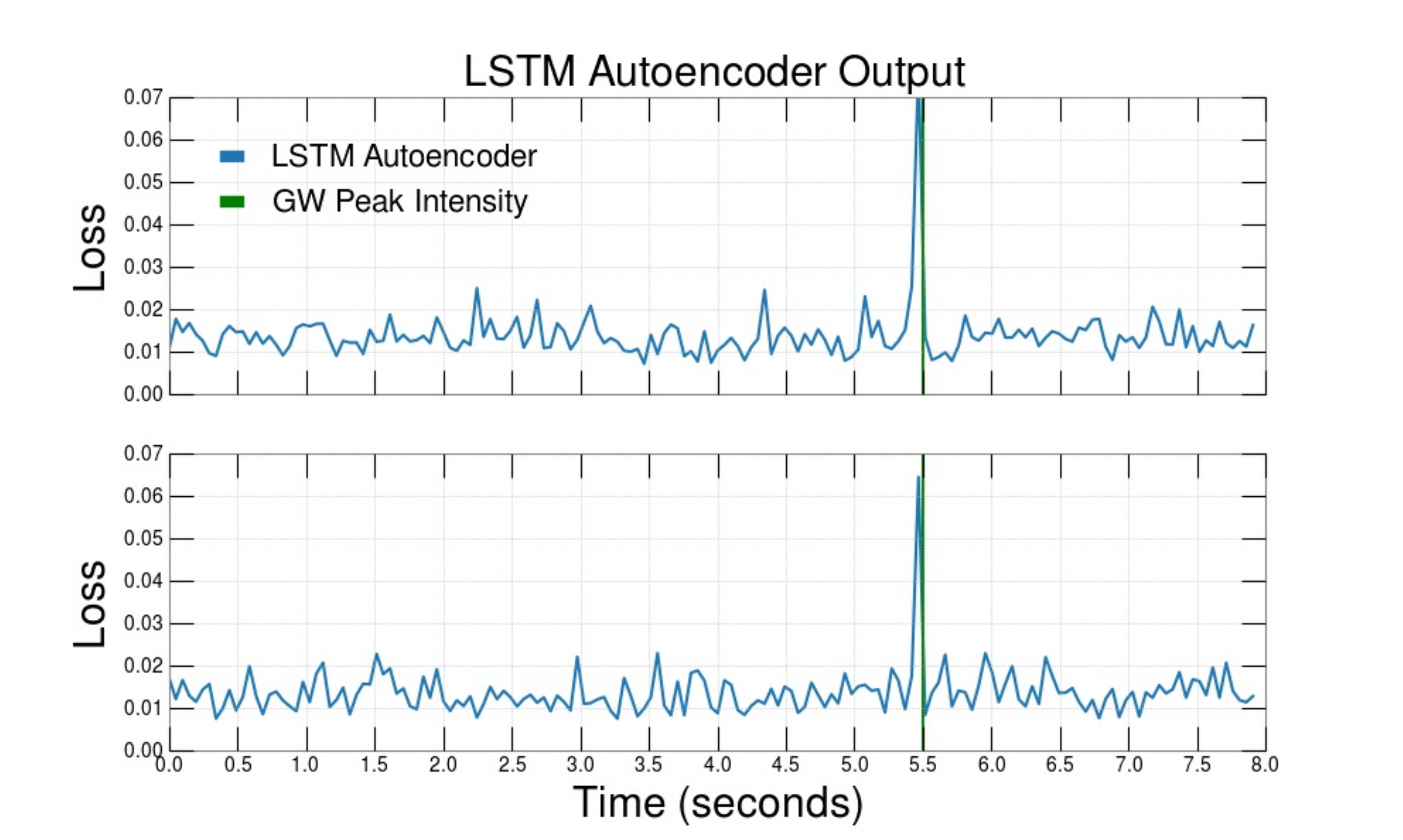}
    \caption{Spike/dip in MSE loss for an LSTM autoencoder over one 8-second event sampled at 2048 Hz (split into 100-timestep inputs) signaling the detection of a gravitational wave for the H1 (top) and L1 (bottom) detectors at an SNR = 18.1. The AE models are never presented with any GW signals (green) during training.  Detection thresholds are set by fixing a FPR, which is done continuously to create a ROC curve in Section~\ref{sec:results}.\label{fig:threshold}}
\end{figure}

The CNN, LSTM, and GRU autoencoders are all trained on unlabeled detector noise data, with no introduction to signal distributions. Doing so, the latent space representation returned by the encoder is exclusively a function of detector noise. As a result, the MSE error is  relatively consistent  during periods with exclusively noise, but it might instantaneously increase when a signal event passes through the autoencoder. An example of such a spike in the MSE loss as a function of time is shown in Fig.~\ref{fig:threshold}. The spike is typically due to the fact that the encoding/decoding sequence learned on noise might not be optimal for a previously unobserved kind of input data. As a result, the distance between the input and the output could be larger for anomalous data, up to generate a spike. Operationally, one could then monitor the MSE value returned by the algorithm, and the detection of a signal could be correlated to the observation of a spike above threshold. 

The performance of the three AE models is assessed comparing their accuracy on benchmark signal samples to that obtained from binary CNN classifiers, trained on the same data and the corresponding labels. Different classifiers are trained for different signal topologies. The classifier architecture is loosely equivalent to that of the CNN encoder from Ref.~\cite{George:2016hay, George:2017pmj}. In particular, it consists of four convolution layers, with 16, 32, 64, and 128 filters respectively, and three fully connected layers with 128, 64, and 2 nodes, respectively. A ReLU activation function was used throughout. Kernel sizes of 4, 4, 4, and 8 were used with dilation rates of 1, 2, 2, and 2 for the convolutional layers and kernel sizes of 2, 2, 4, 4, with a stride of 4 for all the max pooling layers. A sigmoid function is used for the single-node output layer. An LSTM-implementation of the supervised classifier was also attempted but yielded results far worse than the CNN classifier method, so it is not included in this study.  

Two classifiers of this kind are trained, using a dataset of 400,000 samples, consisting an equal fraction of noise events and one of the two classes of signal (BBH and BNS) considered in this study. These classifiers used the same data split (64\% train, 16\% validation, 20\% test) as the autoencoder datasets. The training is performed minimizing a binary cross entropy error loss function on the training sample of Sec.~\ref{sec:data}, using the validation set to optimize the training and the test set to evaluate the model performance. 

The classifier is tested on noise samples, as well as on BBH and BNS events. When tested on the same kind of signal it is trained on, the classifier accuracy is used to estimate the best accuracy that the AE could reach and, consequently, the loss in accuracy due to the use of an unsupervised approach. When testing the classifier on the signal it was not trained on, we can instead compare the generalization property of the autoencoder to that of a supervised algorithm. The two tests provide an assessment of the balance between accuracy and generalization power and demonstrate the complementarity between our approach and a standard template-based method. In practice, one could implement as many supervised algorithms as known GW sources, while using an unsupervised algorithm to be sensitive to unexpected signal sources (and non-coincident signals across multiple interferometers, in cases of glitch detection and data quality monitoring).

\section{Results}
\label{sec:results}

\begin{figure}[t!]
    \centering
    \includegraphics[width=0.49\columnwidth]{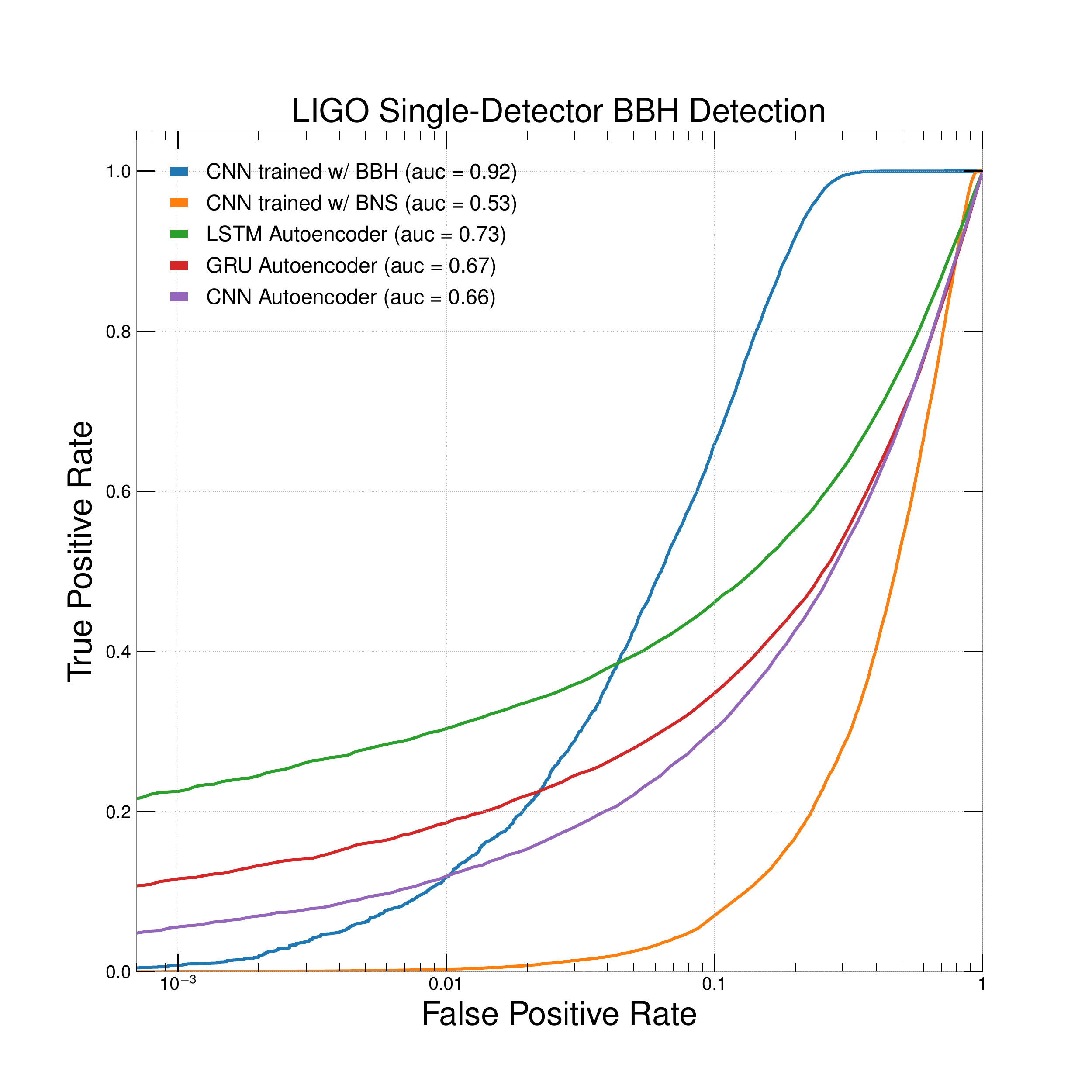} 
    \includegraphics[width=0.49\columnwidth]{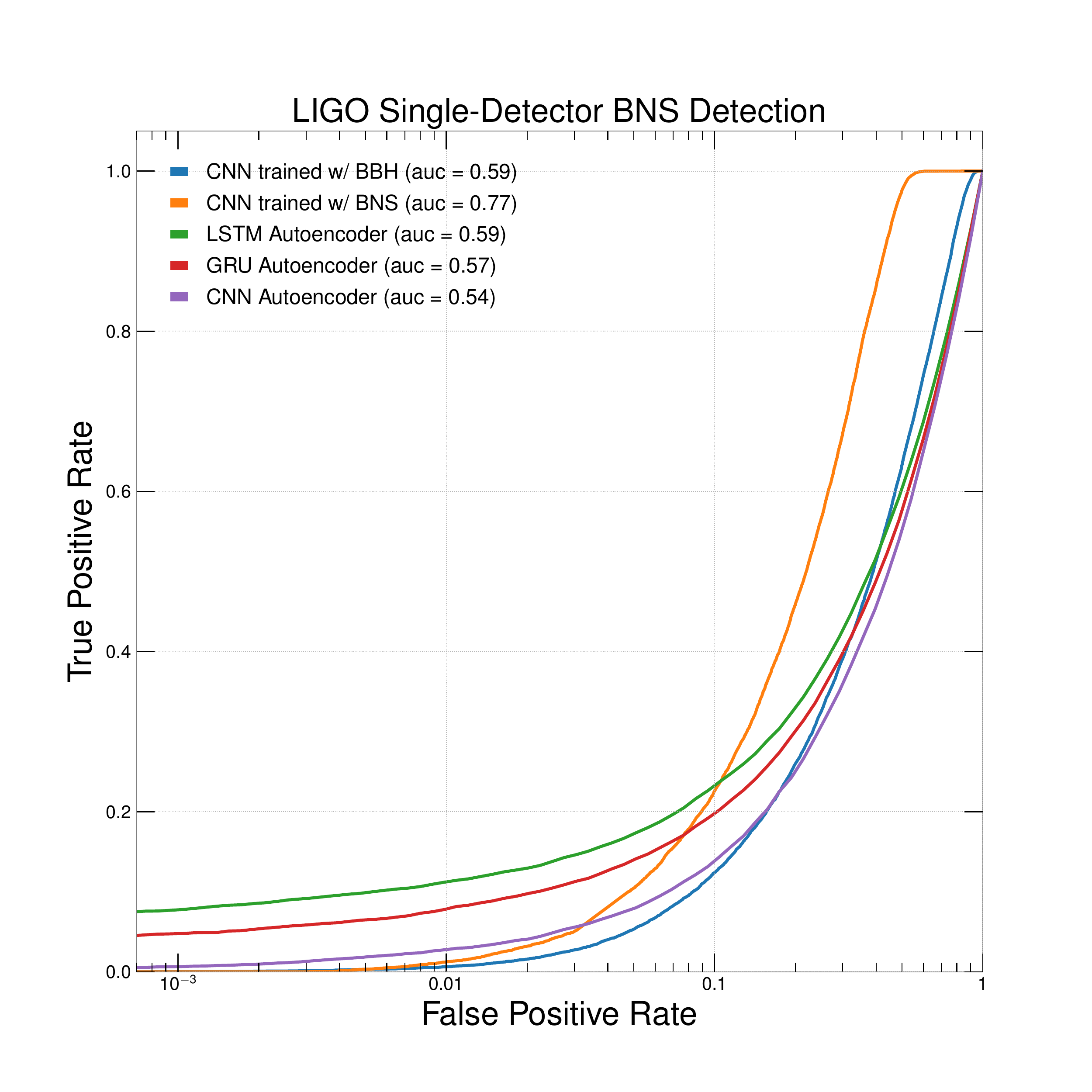}
    \caption{Single-detector ROC curves for the LSTM, GRU, CNN Autoencoders compared to the corresponding CNN supervised architecture. The figure shows the ROC curves for supervised CNNs trained on the BBH and BNS merger data, when the inference is performed on noise and a signal sample of BBH (left) or BNS (right) events. "Area under the ROC Curve" (AUC) measures the two-dimensional area under a ROC curve and quantifies the overall performance of a model. An AUC = 0.50 corresponds to a classifier making random predictions.  \label{fig:ROC}}
\end{figure}

Figure~\ref{fig:ROC} shows the receiver operating characteristic (ROC) curves for three autoencoder architectures (LSTM, GRU, CNN). The curves are obtained considering a single detector, i.e., no coincidence is enforced at this stage. In the left (right) figure, the ROC curves are evaluated on noise and a signal sample of BBH (BNS) merger data. For comparison, the CNN classifiers trained on both datasets are shown. 

As the ROC curves show, the LSTM architecture provides the best accuracy among the AEs. The LSTM AE accuracy is worse than that of the classifier trained on the correct signal hypothesis, but better than that of the classifier trained on the opposing signal hypothesis. Remarkably, the AE outperforms the classifier for a tight enough selection on the network score, i.e., for a low enough target FPR value. The performance comparison is quantified in Table~\ref{tab:FPR_fixed_TPR}, where the false positive rates (FPRs) corresponding to fixed values of the true positive rates (TPRs) are shown. The ROC curves and the valuesf shown in Table~\ref{tab:FPR_fixed_TPR} quantify the trade-off between accuracy and generalization that motivates this study. This makes autoencoders especially useful to potentially discover unexpected GW sources, as well as GW sources that cannot be modeled by traditional simulation techniques.

\begin{table}[ht!]
    \centering
    \caption{True-positive rates for BBH and BNS merger detection at 10\% and 1\% false-positive rates, for autoencoders trained on noise and for binary CNN classifiers trained on BBH and BNS simulations. The autoencoder architecture with the best unsupervised results is marked in bold.
    \label{tab:FPR_fixed_TPR}}
    \begin{tabular}{|c|c|c|c|c|c|c|}
       \hline
       \multicolumn{7}{|c|}{BBH signal vs. noise} \\
       \hline
       FPR & 
       \multicolumn{1}{|c|}{LSTM-AE} &
       \multicolumn{1}{|c|}{GRU-AE} &
       \multicolumn{1}{|c|}{CNN-AE} & &
        \multicolumn{1}{|c|}{BBH-CNN} & 
        \multicolumn{1}{|c|}{BNS-CNN} \\
        \hline
            0.1  & \textbf{46.1}\% & 34.8\% & 30.2\% && 66.0\% & 10.0\% \\ 
            0.01  & \textbf{30.4}\% & 18.6\% & 12.0\% && 11.8\% &  0.3\%\\
        \hline
         \multicolumn{7}{|c|}{BNS signal vs. noise} \\
       \hline
        FPR & 
       \multicolumn{1}{|c|}{LSTM-AE} &
       \multicolumn{1}{|c|}{GRU-AE} &
       \multicolumn{1}{|c|}{CNN-AE} & &
        \multicolumn{1}{|c|}{BBH-CNN} & 
        \multicolumn{1}{|c|}{BNS-CNN} \\ 
        \hline 
            0.1 & \textbf{23.7}\% & 20.2\% & 14.4\% && 12.4\% & 22.5\% \\
            0.01 & \textbf{11.4}\% & 8.16\% & 2.9\% && 0.6\% & 1.6\% \\
       \hline  
    \end{tabular}
\end{table}

This advantage is especially marked with BBHs, which have larger masses and thus shorter signals for the same integrated SNR, leading to a higher instantaneous amplitude at the merger. Thus, autoencoder architectures will likely have an advantage with higher mass-range mergers in the regime where supervised learning models cannot generalize. This is as opposed to BNSs, which have lower mass values and consequently lower amplitude and higher frequency signatures. In this case, the generalization performance stagnates for both of the models, meaning that both models are extracting the same amount of signal out of the events. Still, in the AUC metric the autoencoder models performs equivalently to the supervised algorithm trained on the wrong signal hypothesis and over-performs at low FPR.

\begin{figure}[t!]
    \centering
    \includegraphics[width=0.49\columnwidth]{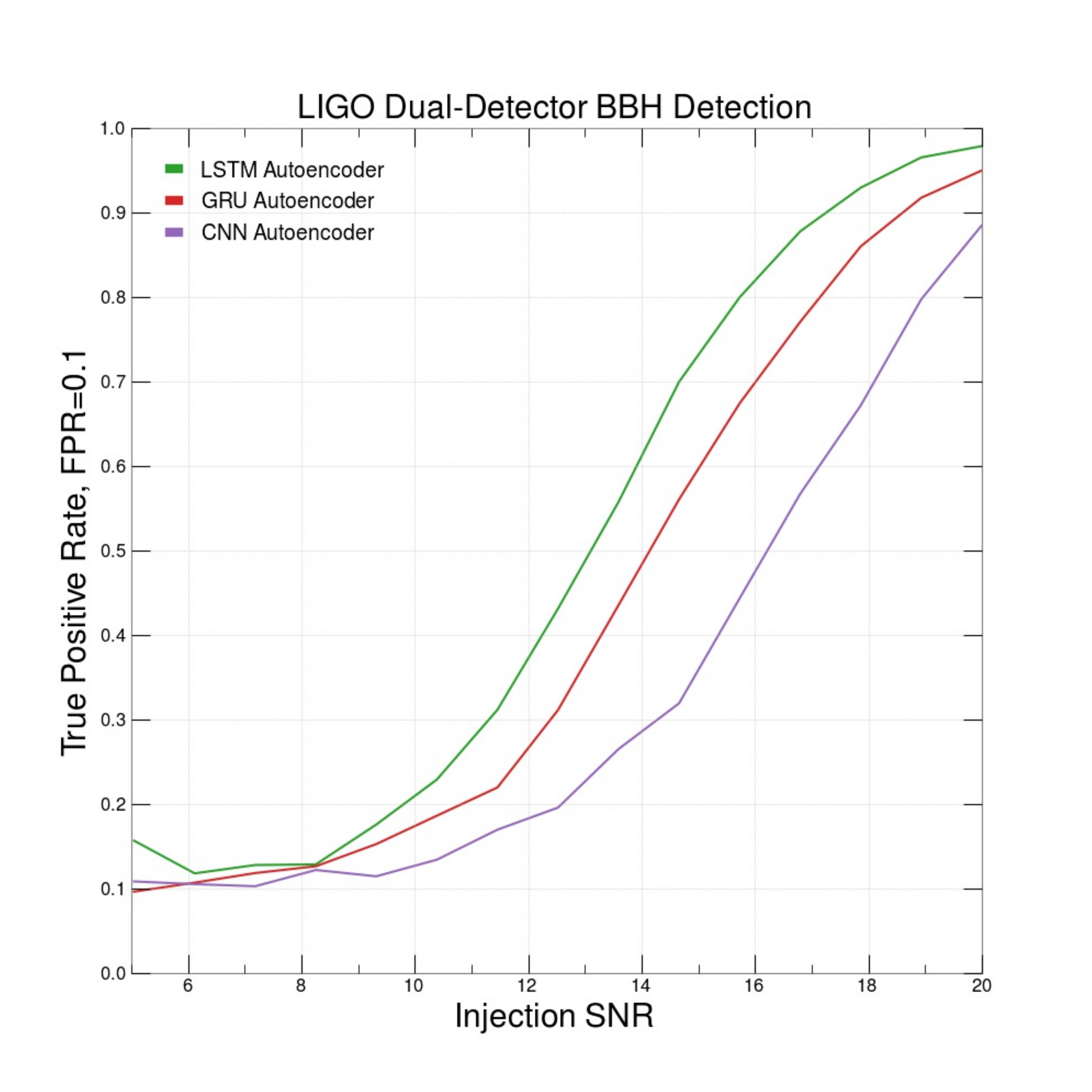}
    \includegraphics[width=0.49\columnwidth]{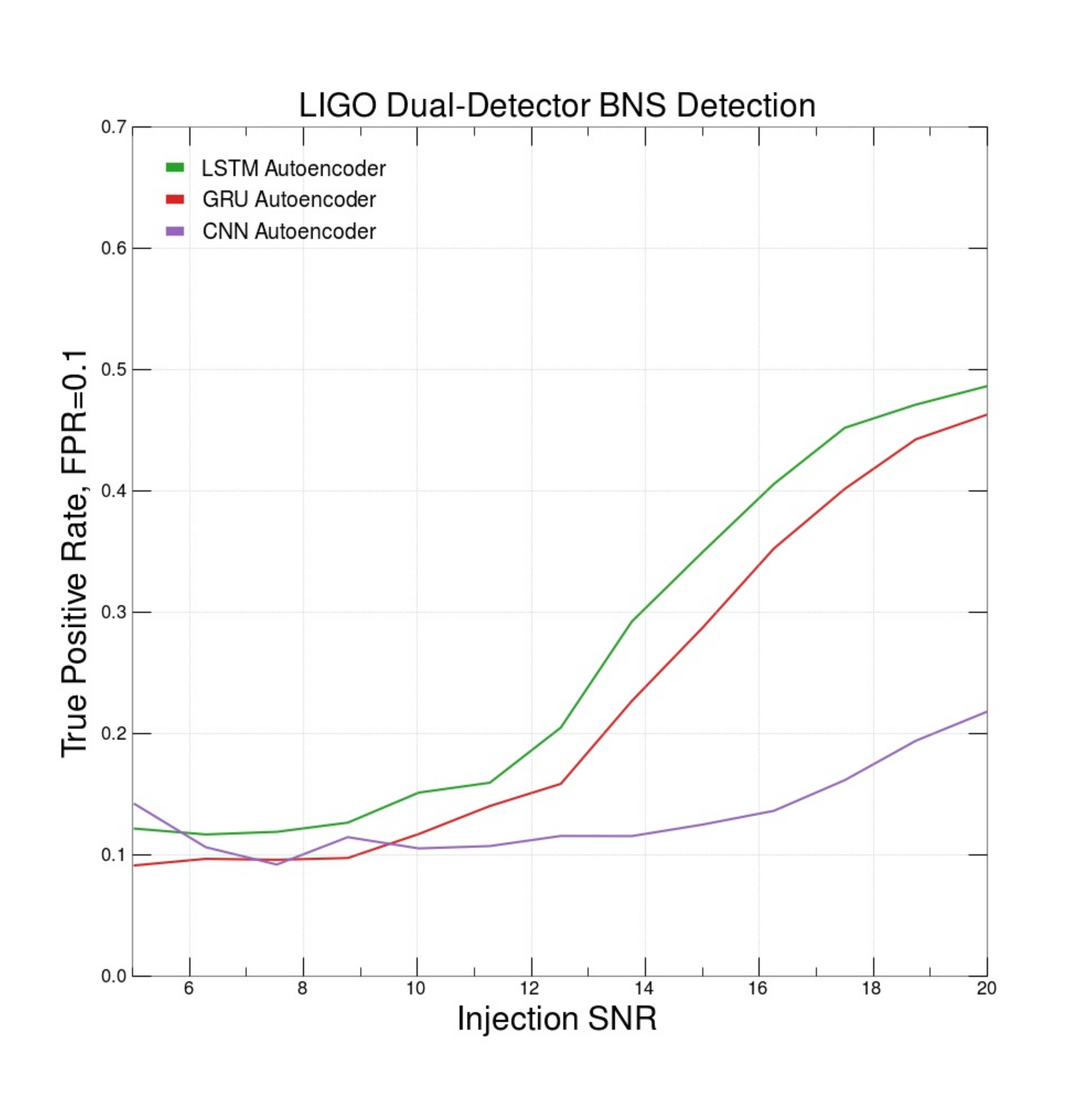}
    \caption{True positive rates for the LSTM Autoencoder at a fixed False Positive Rate (FPR=0.1) with BBH (left) and BNS (right) events at variable Signal-to-noise ratios. \label{fig:SNR}}
\end{figure}

The TPR values quoted on Table~\ref{tab:FPR_fixed_TPR} are obtained averaging across the SNR, which is uniformly distributed in a [5, 20] range. On the other hand, the TPRs of the AE models depend strongly on the SNR value, as shown in Fig.~\ref{fig:SNR} both for BBH and BNS merger events. As shown in the figure, the LSTM AE guarantees better performance across the considered range of SNR values. While the improvement (e.g., with respect to the CNN AE) is roughly constant for BBH events, in the case of BNS events the LSTM AE is particularly better than the other architectures for large SNR values. Overall, this study reinforces the idea that the LSTM AE is the most robust choice among those we considered.

In a realistic exploitation of this algorithm, one would define a threshold above which the data would be called a potential signal. Doing so, one would like to keep the FPR at a manageable rate, while retaining a reasonable TPR value. For instance, an FPR of $10^{-4}$ would correspond to about one false alarm a day, low enough for a post detection assessment of the nature of the anomaly. Similarly, a FPR of $10^{-6}$ would correspond to about one false alarm every three months, low enough for the algorithm to be used in a real-time data processing, e.g., to serve as a trigger for multi-messenger astronomy. 

\begin{figure}[t!]
    \centering
    \includegraphics[width=0.49\columnwidth]{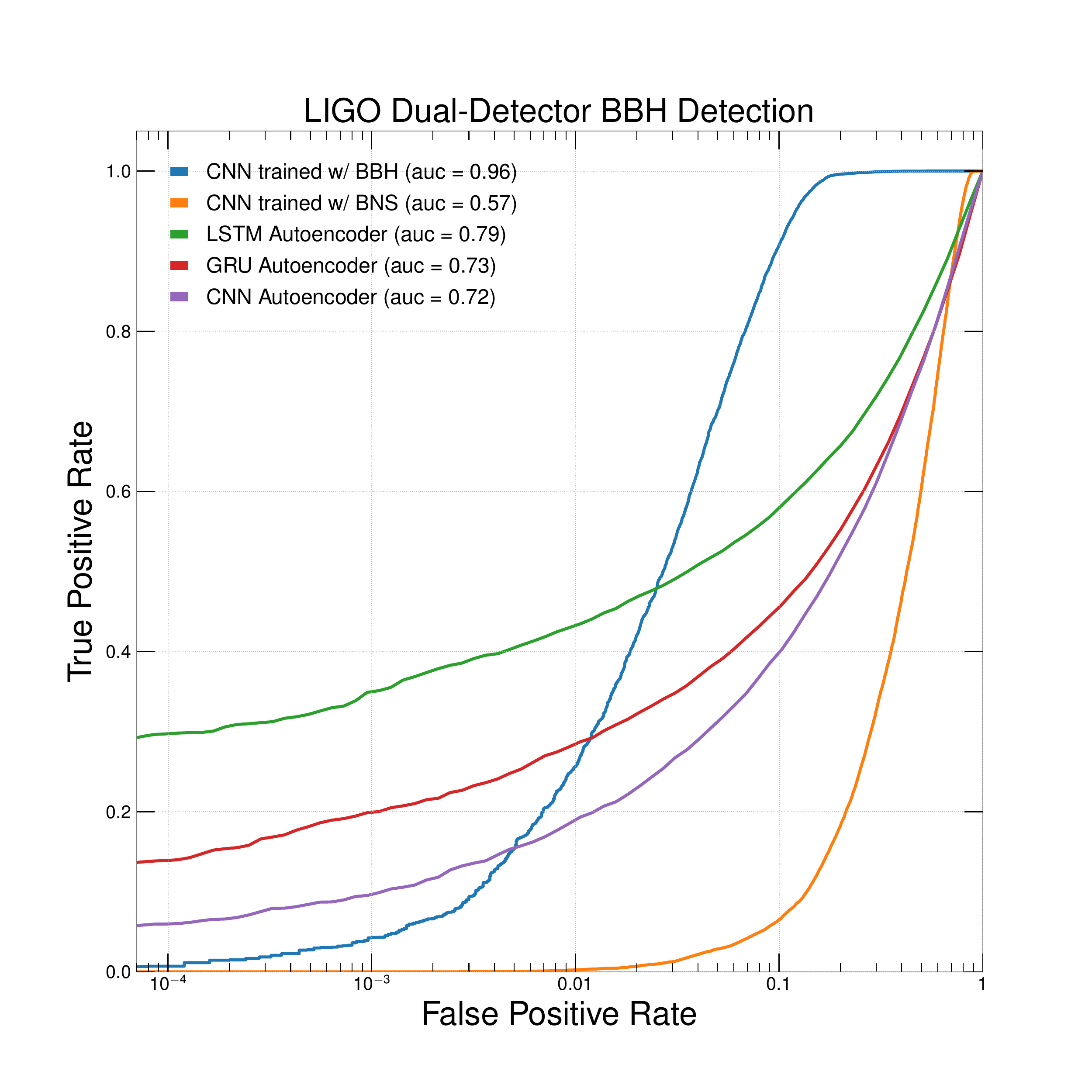}
    \includegraphics[width=0.49\columnwidth]{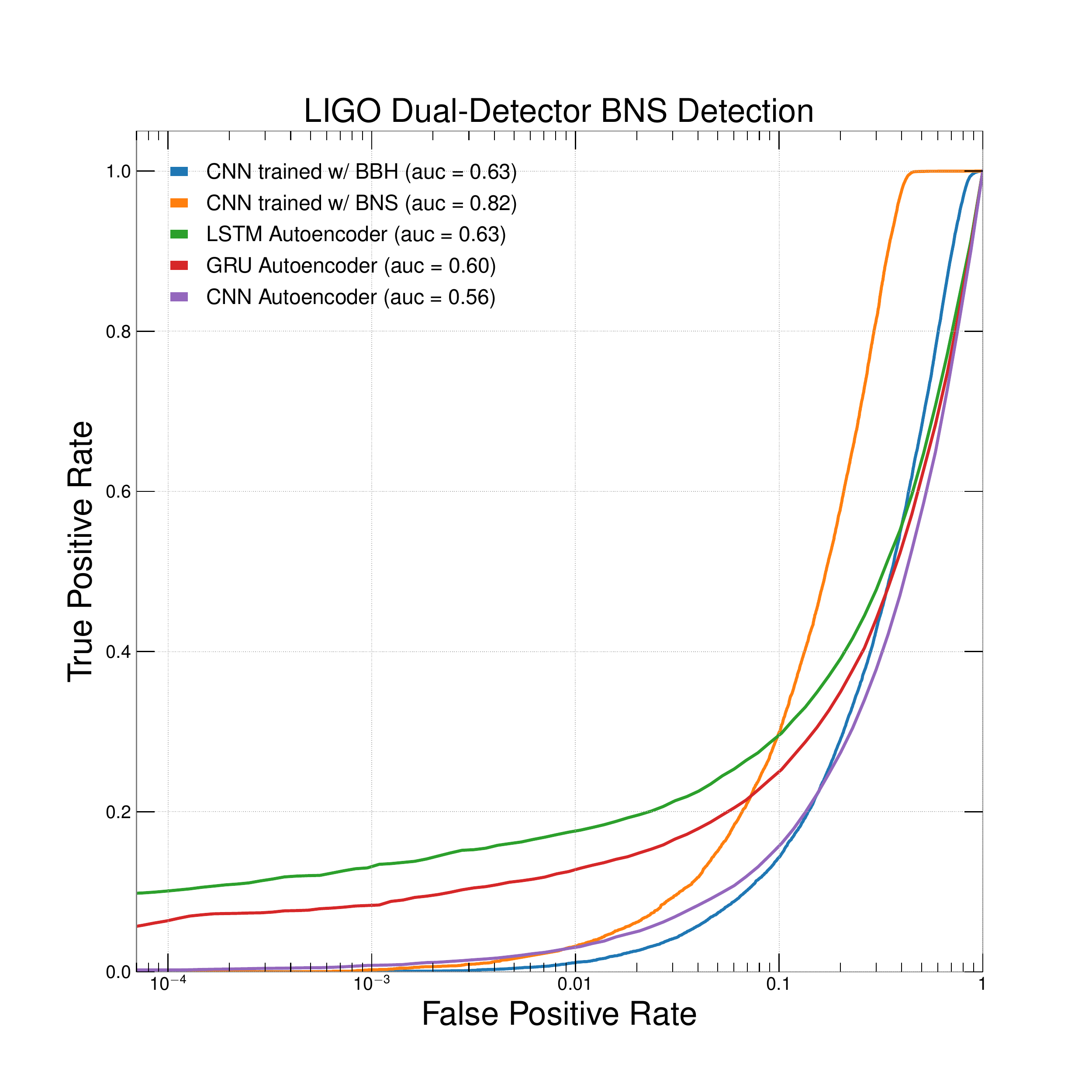}
    \caption{ROC curves for the LSTM, GRU, CNN Autoencoders, obtained by exploiting signal coincidence on two detectors. The loss is computed independently on each detector and the same threshold is applied. The  test is performed on noise and a signal sample of BBH (left) or BNS (right) events. "Area under the ROC Curve" (AUC) measures the two-dimensional area under a ROC curve and quantifies the overall performance of a model. An AUC = 0.50 corresponds to a classifier making random predictions.\label{fig:ROC_coincidence}}
\end{figure}

The key ingredient to reach low FPR values is the exploitation of signal coincidence across multiple detectors. Since the noise across detectors is uncorrelated, single-detector FPR of $\sim 10^{-2}$ (as in Table~\ref{tab:FPR_fixed_TPR}) would give a global FPR of $\sim 10^{-4}$ when two instruments are put in coincidence. Clearly, the presence of uncorrelated noise overlapped to the signal dilutes the correlation of the anomaly across different devices. On the other hand, a certain level of correlation is retained. For instance, we observe a 40\% correlation on the LSTM anomaly score for BBH merger events. For comparison, a 70\% correlation is observed for the CNN classifier. Coincidence can be enforced requiring that two signals above a certain threshold are detected at the same time. Alternatively, one could apply a threshold on the sum of the two losses, with the idea that an MSE loss function is loosely related to the negative log likelihood, so that the sum of the loss would correspond to the negative log of the likelihood products. The former approach has the advantage of requiring the two detectors to communicate only after the anomaly event in a detector is identified. This means that the data throughput to be transmitted can be kept low. On the other hand, the latter approach provides better performance and it is considered here. In this case, one would have to find solutions to mitigate the data throughput and facilitate the communication of the detectors in real time. For instance, one could run the encoder at each experiment site and transmit the compressed data, with the decoding and coincidence check happening off-site. One should keep in mind that the LSTM model can run on a Field Programmable Gate Array within ${\cal O}(100)$ nsec, as demonstrated in Ref.~\cite{que2021accelerating}.

To show this, we consider the case of two detectors and we build a ROC curve requiring a signal above a threshold on the sum of the autoencoder losses. The result is shown in Fig.~\ref{fig:ROC_coincidence}, both for BBH and BNS mergers and quantified in Table~\ref{tab:FPR_fixed_TPR_dual_detector}. Keeping as a target a FPR of $10^{-4}$, one can retain a BBH TPR comparable to that of the $10^{-2}$ FPR of the single-detector threshold (see Table~\ref{tab:FPR_fixed_TPR}), while reducing the FPR by two orders of magnitude. The situation is qualitatively similar for BNS, with some quantitative difference: the two-detector combination comes with a $\sim 20\%$ ($\sim 10\%$) relative reduction of the TPR for a FPR of $10^{-2}$ ($10^{-4}$). Even taking into account, this efficiency loss, there is still a striking advantage in exploiting the coincidence of the signal across detectors to suppress the noise.

\begin{table}[h!]
    \centering
    \caption{True-positive rates for BBH and BNS merger detection at single-detector $10^{-1}$ and $10^{-2}$ false-positive rates corresponding to dual-detector FPRs of $10^{-2}$ and $10^{-4}$, obtained by exploiting signal coincidence in two detectors. The autoencoder architecture with the best unsupervised results is marked in bold.
    \label{tab:FPR_fixed_TPR_dual_detector}}
    \begin{tabular}{|c|c|c|c|c|c|c|}
       \hline
       \multicolumn{7}{|c|}{BBH signal vs. noise} \\
       \hline
       FPR  & 
       \multicolumn{1}{|c|}{LSTM-AE} &
       \multicolumn{1}{|c|}{GRU-AE} &
       \multicolumn{1}{|c|}{CNN-AE} & &
        \multicolumn{1}{|c|}{BBH-CNN} & 
        \multicolumn{1}{|c|}{BNS-CNN} \\
        \hline 
            0.01 & \textbf{43.3}\% & 28.5\% &   18.9\% &  &
            25.6\% &   0.2\% \\ 
            0.0001 & \textbf{29.7}\% &  13.9\% &   6.0\% &  &
            0.7\% &   0.0\% \\
        \hline
         \multicolumn{7}{|c|}{BNS signal vs. noise} \\
       \hline
        FPR  & 
       \multicolumn{1}{|c|}{LSTM-AE} &
       \multicolumn{1}{|c|}{GRU-AE} &
       \multicolumn{1}{|c|}{CNN-AE} & &
        \multicolumn{1}{|c|}{BBH-CNN} & 
        \multicolumn{1}{|c|}{BNS-CNN} \\ 
        \hline 
            0.01 & \textbf{17.6}\% &  12.7\% &   3.1\% &  &
            1.1\% &   3.2\%  \\
            0.0001 & \textbf{10.1}\% &  6.3\% &   0.2\% &  &
            0.0\% &   0.0\% \\
       \hline  
    \end{tabular}
\end{table}

\section{Conclusions}
\label{sec:conclusions}

We presented an unsupervised strategy to detect GW signals from unspecified sources exploiting an AE trained on noise. The AE is trained to compress input data to a low-dimension latent space and reconstruct a  representation of the input from the point in the latent space. The algorithm is optimized using as a loss function a differentiable metric, quantifying the distance between input and output data. Given a trained AE, one could identify anomalous data isolating the outlier data populating the tail of the loss distribution. 

We applied this strategy to a sample of synthetic data from two GW interferometers. We explore different choices for the network architecture and compare the single-detector detection capability to that of a CNN binary classifier, trained on specific signal hypotheses. We show how a recurrent AE provides the best anomaly detection performance on benchmark BBH and BNS merger events. We show the trade-off between accuracy and generalization, when using this unsupervised strategy rather than a supervised approach (here quantified through a CNN binary classifier trained on the same data and the corresponding labels). Additionally, we show that unsupervised AEs perform better than supervised classifiers (trained on the same amount of data) at low enough target FPR values. 

We show how the coincidence of two detectors, both selecting anomalies at an expected FPR$~\sim 10^{-2}$, would retain a TPR of 43.3\% (17.6\%) for BBH (BNS) signals while giving one false alarm a day which can be easily be discarded after a post-detection analysis, e.g., with more traditional GW detection strategies. By comparing this performance to that achieved by other network architectures, we show that our network design, and in particular the use of LSTMs to exploit the time-series nature of the data, represents a progress in terms of anomaly detection capabilities. 

With the same FPR, one could bring the false alarm rate to about once every three month, exploiting the coincidence of three detectors. In this case, the algorithms proposed here could operated in the real-time as part of a trigger system for multi-messenger astronomy, in the spirit of what is discussed in Ref.~\cite{Cerri:2018anq} for real-time data analysis at the Large Hadron Collider. Considering the relatively low computational cost of such an algorithm~\cite{que2021accelerating} and the high impact of a potential signal detection by this algorithm, its implementation for LIGO and VIRGO would be certainly beneficial, despite the fact that the expected detection probability cannot be guaranteed to be high for any signal source.

\section*{Acknowledgments} 

We are grateful to the insight and expertise of Rana Adhikari, Hang Yu, and Erik Katsavounidis from the LIGO collaboration and Elena Cuoco from the VIRGO collaboration, who guided us on a field of research which is not our own.

Part of this work was conducted at "\textit{iBanks}", the AI GPU cluster at Caltech. We acknowledge NVIDIA, SuperMicro and the Kavli Foundation for their support of "\textit{iBanks}". 

This work was carried on as part of the 2020 CERN OpenLab Summer Student program, which was carried on in remote mode due to the COVID pandemic. 

M. P. is  supported by the European Research Council (ERC) under the European Union's Horizon 2020 research and innovation program (Grant Agreement No. 772369).

E. M. is supported by the Institute for Research and Innovation in Software for High Energy Physics (IRIS-HEP) through a fellowship in Innovative Algorithms. 

This work is partially supported by the U.S. DOE, Office of Science, Office of High Energy Physics under Award No. DE-SC0011925, DE-SC0019227 and DE-AC02-07CH11359.

\section*{References}
\bibliographystyle{unsrt}
\bibliography{bib}


\end{document}